%% file: 990123-paper.tex

\def\singlespace{\baselineskip 12pt \lineskip 1pt \parskip 2pt plus 1 pt}


\singlespace
\magnification=\magstep1


\raggedbottom

\def\refto#1{$^{#1}$}           
\def\ref#1{ref.~#1}                     
\def\Ref#1{#1}                          
\gdef\refis#1{\item{#1.\ }}                     
\def\beginparmode{\endmode
  \begingroup \def\endmode{\par\endgroup}}
\let\endmode=\par
\def\body{\beginparmode}
\def\head#1{                    
  \goodbreak\vskip 0.5truein    
  {\centerline{\bf{#1}}\par}
   \nobreak\vskip 0.25truein\nobreak}
\def\references                 
  {\head{References}            
   \beginparmode
   \frenchspacing \parindent=0pt \leftskip=1truecm
   \parskip=8pt plus 3pt \everypar{\hangindent=\parindent}}
\def\endreferences{\body}

\font\lgh=cmbx10 scaled \magstep2

\def\hb{\hfil\break}

\def\ale{\mathrel{\hbox{\rlap{\hbox{\lower4pt\hbox{$\sim$}}}\hbox{$<$}}}}
\def\age{\mathrel{\hbox{\rlap{\hbox{\lower4pt\hbox{$\sim$}}}\hbox{$>$}}}}

\input reforder.tex

\input citmac.tex


\def\Discovery-Image{1}
\def\Light-Curves{2}
\def\Spectrum-Figure{3}

\def\TableMagnitudes{1}
\def\Table-Lines{2}


\hrule
\bigskip
\line{\lgh The afterglow, the redshift, and the extreme \hb}
\line{\lgh energetics of the $\gamma$-ray burst 990123 \hb}
\bigskip

\def\Palomar  {$^1$}
\def\IPAC     {$^2$}
\def\TAPIR    {$^3$}
\def\UCSC     {$^4$}
\def\DTM      {$^5$}
\def\Kapteyn  {$^6$}
\def\NRAO     {$^7$}
\def\Carnegie {$^8$}
\def\UCLA     {$^9$}
\def\GSFC     {$^{10}$}
\def\UPenn    {$^{11}$}
\def\NAOJ     {$^{12}$}
\def\Columbia {$^{13}$}
\def\CARA     {$^{14}$}
\def\IAS      {$^{15}$}

\line{S. R. Kulkarni\Palomar,           
      S. G. Djorgovski\Palomar,         
      S. C. Odewahn\Palomar,            
      J. S. Bloom\Palomar,              
      R. R. Gal\Palomar,                
\hb}
\line{C. D. Koresko\Palomar,            
      F. A. Harrison\Palomar,           
      L. M. Lubin\Palomar,              
      L.    Armus\IPAC,                 
      R.    Sari\TAPIR,                 
\hb}
\line{G. D. Illingworth\UCSC,           
      D. D. Kelson\DTM,                 
      D. K. Magee\UCSC,                 
      P. G. van Dokkum\Kapteyn,         
\hb}
\line{D. A. Frail\NRAO,                 
      J. S. Mulchaey\Carnegie,          %
      M. A. Malkan\UCLA,                
      I. S. McLean\UCLA,                
      H. I. Teplitz\GSFC,               
\hb} 
\line{D.    Koerner\UPenn,              
      D.    Kirkpatrick\IPAC,            
      N.    Kobayashi\NAOJ,             
      I. A. Yadigaroglu\Columbia,       
\hb}
\line{J.    Halpern\Columbia,           
      T.    Piran\Columbia,             
      R. W. Goodrich\CARA,              
      F.    Chaffee\CARA,               
      M.    Feroci\IAS,                 
      E.    Costa\IAS                   
\hb}

\bigskip

\line{\Palomar Palomar Observatory 105-24, Caltech, Pasadena, CA 91125, USA\hb}

\line{\IPAC Infrared Processing \&\ Analysis Center, Caltech, 
Pasadena, CA 91125, USA\hb}

\line{\TAPIR Theoretical Astrophysics 130-33, Caltech, Pasadena CA 91125, USA
\hb}

\line{\UCSC Lick Observatory, Univ. of California, 
            Santa Cruz, CA 95064, USA \hb}

\line{\DTM Dept. of Terrestrial Magnetism, 
           Carnegie Institution of Washington,\hb}
\line{\ \ 5241 Broad Branch Rd., NW, Washington DC 20015, USA \hb}

\line{\Kapteyn Kapteyn Astron. Institute, P.O. Box 800,
            NL-9700 AV, Groningen, The Netherlands \hb}

\line{\NRAO National Radio Astronomy Observatory, P. O. Box O,
       Socorro, NM 87801, USA\hb}

\line{\Carnegie Observatories of the Carnegie Institution, 813 Santa
Barbara Street,\hb}

\line{\ \ Pasadena, CA 91101, USA \hb}

\line{\UCLA Dept. of Physics \&\ Astronomy, Univ. of California,
Los Angeles, CA 90095-1562, USA \hb}

\line{\GSFC Goddard Space Flight Center, Code 681, Greenbelt, MD 20771,
USA \hb} 

\line{\UPenn Univ. of Pennsylvania, 4N14 DRL, 209 S. 33rd St.,
             Philadelphia, PA 19104-6396, USA \hb}

\line{\NAOJ Subaru Telescope, National Astronomical Observatory of Japan,\hb}
\line{\ \ \ 650 N. A'ohoku Place, Hilo, HI 96720, USA \hb}

\line{\Columbia Dept. of Astronomy, Columbia Univ., 538 W. 120th St., New York,
NY 10027, USA\hb}

\line{\CARA W. M. Keck Observatory, 65-1120 Mamalahoa Highway, 
            Kamuela, HI 96743, USA \hb}

\line{\IAS Istituto di Astrofisica Spaziale, CNR,  
           via Fosso del Cavaliere, Roma I-00133, Italy\hb}

\medskip
\hrule
\bigskip

\noindent{\it  This manuscript was submitted to Nature on 11 February
1999 and resubmitted on 27 February 1999 with revisions in response to
comments from the referees. We are making this m.s. available on
astroph given the intense interest in GRB 990123. You are free to refer
to this paper in your own paper. However, we do place restrictions on
any dissemination in the popular media. The article is under embargo
until it is published. For further enquiries, please contact Shri
Kulkarni (srk@astro.caltech.edu) or Fiona Harrison
(fiona@srl.caltech.edu).

A companion paper to this by Bloom et al. can also be found on astroph.
(astr-ph/9902182) with links to HST images of the host of GRB 990123.}
\vfill\eject

\noindent{\bf Afterglow, or long-lived emission, has now been detected
from about a dozen well-positioned gamma-ray bursts.  Distance
determinations made by measuring optical emission lines from the host
galaxy, or absorption lines in the afterglow spectrum, place the burst
sources at significant cosmological distances, with redshifts ranging
from $\sim$1--3.  The energy required to produce the bright
gamma-ray flashes is enormous: up to $\sim 10^{53}$ erg or 10\%
of the rest mass energy of a neutron star, if the emission is isotropic.
Here we present the discovery of
the optical afterglow and the redshift of GRB 990123, the brightest
well-localized GRB to date.  With our measured redshift of $\geq 1.6$,
the inferred isotropic energy release exceeds the rest mass of
a neutron star thereby challenging
current theoretical models for the origin of GRBs.  We argue that the
optical and IR afterglow measurements reported here may provide the
first observational evidence of beaming in a GRB, thereby reducing the
required energetics to a level where stellar death models are still
tenable.  }

\bigskip
\bigskip


Almost thirty years after their discovery\refto{Klebesadel73}, the
distance scale to gamma-ray bursts (GRBs) was finally unambiguously
determined just two years ago\refto{Metzger97}.  Thanks to precise and
prompt GRB localizations by the Italian-Dutch satellite,
BeppoSAX\refto{Boella97} and the All Sky Monitor\refto{Levine96}
aboard the Rossi X-ray Timing Explorer, 
afterglow emission spanning the wavelength range from X-ray to radio has
now been detected in about a dozen events\refto{Costa97,jvP97,Frail97}.
This has provided redshift measurements for four bursts: one through
the detection of optical absorption lines\refto{Metzger97} and three
by identifying emission features in the host
galaxies\refto{Kulkarni98,Djorgovski98a,gcn189}.  It is now assumed
that a substantial fraction of all GRBs occur at significant
cosmological distances ($z \sim$ 1--3).

\refis{Klebesadel73}
        Klebesadel, R. W., Strong, I. B., \&\ Olson, R. A.
        Observations of gamma-ray bursts of cosmic origin.  
        {\it Astrophys. J.} {\bf 182}, L85--L88 (1973).

\refis{Metzger97}   
        Metzger, M. R., Djorgovski, S. G., Kulkarni, S. R.,
        Steidel, C. C., Adelberger, K. L., Frail, D. A.,
        Costa, E. \&\ Frontera, F. 
        Spectral Constraints on the redshift of the optical
        counterpart to the gamma-ray burst of May 8, 1997.
        {\it Nature} {\bf 387}, 878--879 (1997).

\refis{Boella97}
	Boella, G.  et al.
	BeppoSAX, the wide band mission for x-ray astronomy.
	{\it Astron. Astrophys. Suppl. Ser.} {\bf 122}, 299--399 (1997).

\refis{Levine96}
	Levine, A. M., Bradt, H., Cui, W., Jernigan, J. G., 
	Morgan, E. H., Remillard, R., Shirey, R. E. \&\ Smith, D. A.
	First Results from the All-Sky Monitor on the Rossi X-ray
	Timing Explorer.
	{\it Astrophys. J.} {\bf 469}, L33-L36 (1996).

\refis{Costa97} Costa, E. et al.
            Discovery of an X-ray afterglow associated with the gamma-ray
            burst of 28 February 1997.
            {\it Nature} {\bf 387}, 783--785 (1997).

\refis{jvP97}
	van Paradijs, J.  et al.
	Transient optical emission from the error box of the $\gamma$-ray burst
	of 28 February 1997.
	{\it Nature} {\bf 386} 686--689 (1997).

\refis{Frail97}
	Frail, D. A., Kulkarni, S. R., Nicastro, L., Feroci, M. 
	\&\ Taylor, G.  B.
	The radio afterglow from the gamma-ray burst of 8 May 1997.
	{\it Nature} {\bf 389}, 261--263 (1997).

\refis{Kulkarni98}
        Kulkarni, S.~R. et~al.
        Identification of a host galaxy at redshift $z$=3.42 for the
        $\gamma$-ray burst of 14 Dec 1997.
        {\it Nature} {\bf 393}, 35--39 (1998).

\refis{Djorgovski98a}
        Djorgovski, S. G., Kulkarni, S. R., Bloom, 
        J. S., Goodrich, R., Frail, D. A., Piro, L., \&\ Palazzi, E.
        Spectroscopy of the Host Galaxy of the Gamma--Ray Burst 980703.
        Astrophys. J. {\bf 508}, L17--L20 (1998).

\refis{gcn189}
        Djorgovski, S. G., Kulkarni, S. R., Bloom, J. S., Frail, D. A.,
        Chaffee, F. \&\ Goodrich, R. 
        GRB 980613: Spectroscopy of the host galaxy.
        {\it G.C.N.} {\bf 189}, (1999).

The large inferred distances imply staggering energetics. In
particular,  our measured redshift of $z=3.14$ for GRB 971214 (ref.
\Ref{Kulkarni98}) led us to an inferred isotropic energy loss of
$3\times 10^{53}$ erg, or $0.1M_\odot c^2$. At that time, this result
was seen to challenge most proposed GRB models.  The energy estimate
can be reduced by invoking non-spherical emitting surface geometry,
e.g. jets. However,  to date, there is no firm observational evidence
for jets in GRBs. Indeed, on statistical grounds,
Grindlay\refto{Grindlay99} and Greiner et al.\refto{Greiner98}
constrain the ``beaming factor'' (the reduction in inferred energy
release relative to that for the spherical case) to be less than 0.1.

\refis{Grindlay99} Grindlay, J. 
	Fast X-ray Transients and Gamma-ray Bursts: Constraints on
	Beaming.
	{\it Astrophys. J.} {\bf 510}, 710--714 (1999).

\refis{Greiner98} Greiner, J., Voges, W., Boller, T. \&\ Hartmann, D.
	Search for GRB afterglows in the ROSAT all-sky survey.
	{\it Astron. Astrophys.,} submitted (1999).

BeppoSAX ushered in 1999 with the discovery of
GRB~990123\refto{Heise99}, the brightest GRB seen by BeppoSAX to date.
It is in the top 0.3\% of all bursts, if ranked by the observed
fluence\refto{gcn224}.  If the inferred energetics of GRB 971214 
strained theoretical models, GRB~990123, with a minimum redshift of 1.6
and an inferred isotropic energy release of $1.9M_\odot c^2$, takes
them to the breaking point.  In this paper, we present optical and
infrared observations of GRB~990123, which both establish the distance
to this luminous event, and through the observed break in the 
decay may provide the first observational evidence yet of beaming in a
GRB.

\refis{Heise99}
	Heise, J. et al. 
	A very energetic $\gamma$-ray burst on January 23 1999 and
	its X-ray afterglow.
	{\it Nature}, submitted (1999).

\refis{gcn224}
        Kippen, R. M.
        GRB 990123: BATSE Observations
                {\it G.C.N.} {\bf 224} (1999)

\bigskip
\noindent{\bf The optical and the infrared afterglow}
\medskip

Much of our information about the parameters of GRB explosions
(energy, ambient gas density, dynamics) has been obtained from radio
and optical afterglow studies. These returns have motivated observers
to carry out long-term multi-wavelength studies of the afterglow.

We initiated an optical and IR follow-up program for GRB~990123 on
23.577 January 1999 UT, 3.7 hr after the event, in response to the
BeppoSAX preliminary localization\refto{gcn199}.  We used a charge
coupled device (CCD) at the Palomar 60-inch telescope to image the
field of the 
BeppoSAX localization. We identified a new source by comparing the first CCD
image with an image of the field obtained from the Digital Palomar
Observatory Sky Survey II (DPOSS), and concluded that this
was the optical afterglow of GRB 990123. We promptly reported the
discovery to the GRB Coordinates Network\refto{GCN} (GCN)
and the Central
Bureau of Astronomical  Telegrams. 
The optical source lies within the tighter BeppoSAX
localization\refto{Heise99} obtained a few hours later (see Figure
\Discovery-Image).

\refis{gcn199}
        Piro, L.
        GRB990123, BeppoSAX WFC detection and NFI planned follow-up.
        {\it G.C.N.} {\bf 199} (1999).

\refis{GCN}
	The GRB Coordinates Network.
http://gcn.gsfc.nasa.gov/gcn/

Continued observations at Palomar, Keck and other observatories showed
that the source exhibited the characteristic fading behavior seen in
other GRB optical afterglows.  The association of the optical transient
(OT) with GRB 990123 is therefore secure. Table \TableMagnitudes\
summarizes the photometric observations of the OT.  Included in the
table is a single observation\refto{gcn254} obtained from the Hubble
Space Telescope (HST); the analysis of the HST data are reported
elsewhere\refto{Bloom99}.  Figure \Light-Curves\ shows the OT light
curve in various bands (primarily Gunn $r$ and $K$) using the photometric
data shown in Table \TableMagnitudes.  

\refis{gcn254}
	Beckwith, S.
	For Immediate Posting to GCN -- HST Data GRB 990123 Available.
        {\it G.C.N.} {\bf 254} (1999).

\refis{Bloom99}
	Bloom, J. S. et al. 
	The host galaxy of GRB 990123.
	http://xxx.lanl.gov, astro-ph/9902182 (1999).

A considerable body of
literature\refto{Katz94,MR97,Vietri97,Waxman97a} has been developed to
explain the afterglow phenomenon. Briefly, afterglow can be understood
as arising from synchrotron emission from particles shocked by the
explosive debris sweeping up ambient medium.  Assuming that the
electrons behind the shock are accelerated to a power-law differential
energy distribution with index $-p$, the simplest afterglow model
(e.g. ref. \Ref{Sari98}) predicts that the afterglow flux,
$f_\nu(t)\propto t^{\alpha}\nu^{\beta}$ where $f_{\nu}(t)$ is the flux
at frequency $\nu$ and $t$ is the time measured with respect to the
epoch of the GRB.  Interpretation of the afterglow data in the
framework of these models can, in principle, yield many interesting
parameters of the GRB explosion (e.g. ref.
\Ref{Wijers98,Ramaprakash98}).

\refis{Katz94}  Katz, J.I.
                Low-frequency spectra of gamma-ray bursts.
                {\it Astrophys. J.} {\bf 432}, L110--113 (1994).

\refis{MR97}   M\'esz\'aros, P., \& Rees, M.J.
               Optical and Long-Wavelength Afterglow from Gamma-Ray Bursts.
               {\it Astrophys. J.} {\bf 476}, 232--240, (1997).

\refis{Vietri97} Vietri, M.
               The Soft X-Ray Afterglow of Gamma-Ray Bursts, A Stringent
               Test for the Fireball Model.
               {\it Astrophys. J.} {\bf 478}, L9--12. (1997).

\refis{Waxman97a}   
               Waxman, E.
               Gamma-ray-burst afterglow: supporting the cosmological
               fireball model, constraining parameters, and making
               prediction.
               {\it Astrophys. J.} {\bf 485}, L5--L8 (1997).

\refis{Sari98}
        Sari, R., Piran, T., Narayan, R.
        Spectra and Light Curves of Gamma-Ray Burst Afterglows.
        {\it  Astrophys.~J.} {\bf 497}, L17--L20 (1998).

\refis{Wijers98} Wijers, R. A. M. J. \&\ Galama, T. J. 
	Physical parameters of GRB 970508 and GRB 971214 from
	their afterglow synchrotron emission.
	http://xxx.lanl.gov, astro-ph/9806175 (1998).
	
\refis{Ramaprakash98} Ramaprakash, A. N. et al. 
	The energetic afterglow of the gamma-ray burst of 14 December 1997.
	{\it Nature} {\bf 393}, 43--46 (1998).

As the OT faded, the $K$ band image of January 28 showed that the host
galaxy was only a fraction of an arcsecond displaced with respect to
the OT. A proper analysis of the decay of the OT thus requires that we
correctly assess the contribution of the host galaxy to the flux of
the OT.  The adopted values are $K_{\rm host}= 0.9 \pm 0.3\,\mu$Jy and
$r_{\rm host}= 0.58\pm 0.04\,\mu$Jy; see legend to Figure
\Light-Curves for additional details.  A discussion of the host galaxy
can be found in Bloom et al.\refto{Bloom99}.

In Figure \Light-Curves we present the light curve of the OT after
subtracting the contribution from the host galaxy and correcting for
Galactic extinction. No single power law can fit the $r$-band curve.
We
fit two power laws with a break at epoch $t_{\rm break}$ (see Figure
\Light-Curves). The best fit parameters are $\alpha_{1r}=-1.10\pm 0.03$,
$t_{\rm break}=2.04 \pm 0.46$ d and $\alpha_{2r}= -1.65\pm 0.06$.  In
contrast, the $K$ band data are well fitted by a single power law,
$\alpha_{K}=-1.10 \pm 0.11$. 
We now discuss the behaviour of the afterglow
in the first few days.

The value of $\alpha_{1r}$ and $\alpha_K$  are similar to those
measured in other afterglows (e.g. refs.  \Ref{Sokolov98,Groot97,
Kulkarni98}).  In the 2--10 keV band the afterglow was observed by
a variety of instruments aboard
BeppoSAX starting 6 hr after the burst\refto{Heise99}. From the
analysis of these data, $\alpha_X=-1.44 \pm 0.07$.  In the framework of
the afterglow models, the power law decay index \hbox{$\alpha =
3(1 - p)/4$} or \hbox{$\alpha = (2 -3p)/4$}, depending on whether the
electrons are cooling on a timescale that is slower or faster than the
age of the shock.  A value of $p=2.44$ is consistent with
$\alpha_{1r}$, $\alpha_K$ and $\alpha_X$; this would require that the
X-ray emitting electrons are in the fast cooling regime whereas the
electrons emitting photons in the $r$ and the $K$ bands are in
the slow cooling regime.

\refis{Sokolov98}
	Sokolov, V. V., Kopylov, A. I., Zharikov, S. V., Feroci, M.,
	Nicastro, L. \&\ Palazzi, E. 
	BVR$_C$I$_C$ photometry of GRB 970508 optical remnant:
        May-August 1997.
	{\it Astron. Astrophys.} {\bf 334}, 117--123 (1998).

\refis{Groot97}
	Groot, P. et al.
	The decay of optical emission from the gamma-ray burst GRB 970228.
	{\it Nature} {\bf 387} 479--481 (1997)

As briefly summarized above, afterglow models predict that the
spectrum of the OT should also be a power law and with the spectral
index $\beta$ having a specific relation to $\alpha$.  After
correcting for Galactic extinction, one day after the burst we find
that $\beta_{rK}=-0.8\pm 0.1$. This value is what is expected when the
electrons emitting optical and IR photons are in the slow cooling
regime.  In contrast, comparing the X-ray flux six hours after the
burst\refto{Heise99} with the interpolated $r$-band flux (from Figure
\Light-Curves) we find $\beta_{rX}=-0.54$.  Such a large value for
$\beta$ is inconsistent with the simple afterglow model.  (Supporting
this flat spectrum is the observation that the $B$-band fluxes, although
limited, are comparable to the $r$-band fluxes; see Figure
\Light-Curves.)  Selective extinction by intervening dust in the host
galaxy is an
unlikely explanation given the strong detection of the afterglow
in the $B$ band (Table 1) and in the $U$ band\refto{gcn214}.

\refis{gcn214}
        Falco, E., Petry, C., Impey, C., Koekemoer, A., Rhoads, J.
        GRB 990123 optical observations.
        {\it G.C.N.} {\bf 214} (1999).

We draw attention to the fact that such a discrepancy between X-ray and
optical fluxes is not uncommon and, in our opinion, a similiar
discrepancy exists for GRB 971214 and GRB 980703 as well. This topic is
worthy of further study since it is indicative of a fundamental
violation of at least one basic assumption of the simple afterglow
model (spherical geometry, impulsive energy release and only one
dominant emission mechanism -- synchrotron emission) Indeed, later in
this article, we show that the later behaviour of the afterglow is also
not consistent with the simplest afterglow model.

\bigskip
\noindent{\bf Determining the redshift}
\medskip

We obtained spectra of the optical transient source with the Low
Resolution Imaging Spectrograph\refto{Oke1995} on the Keck II telescope
on the night of January 23, 1999.  These observations consisted of
three spectra, each of about 600-s duration. A quick analysis showed
absorption features, indicating that the OT must be at or beyond a
redshift of 1.61.  A subsequent detailed data reduction yielded the
spectrum displayed in Figure \Spectrum-Figure.

\refis{Oke1995} Oke, J. B., Cohen, J. L., Carr, M., Cromer, J.,
              Dingizian, A., Harris, F. H., Labrecque, S., Lucinio, R.
              Schaal, W., Epps, H. \&\ Miller, J.
              The Keck Low-Resolution Imaging Spectrometer.
              {\it Publ. Astr. Soc. Pacific} {\bf 107}, 375-385 (1995).

The OT spectrum is marked by prominent absorption lines whose suggested
identifications are listed in Table \Table-Lines, and marked in Figure
\Spectrum-Figure. Similar lines are routinely seen in quasar spectra,
and are thought to arise from absorption in intergalactic clouds.
Accepting the identifications, we note that all the absorption features
can be attributed to a single intergalactic cloud at a redshift
$z_{abs}=1.6004\pm 0.0008$. The uncertainty in the redshift has
approximately equal contributions from random and systematic errors.
The absorption is relatively strong, suggesting the absorbing system
has a high column density of gas. This is only the second time that an
absorption redshift has been determined for a GRB OT (the other case
being GRB 970508 [ref.  \Ref{Metzger97}], and possibly also GRB 980703
[ref.  \Ref{Djorgovski98a}]), thus placing the OT unambiguously at a
cosmological distance.

Although the redshift,  $z_{abs} = 1.6$, is a lower limit, it seems
likely that the absorption originates in the GRB host galaxy.  From the
absence of the Ly$\alpha$ forest in our spectrum, we can place a firm
upper limit to the redshift of $z < 2.9$.  The absence of the reported
Ly$\alpha$ forest in the spectrum\refto{gcn219} obtained from the
Nordic Optical Telescope (NOT)\refto{gcn219} would further lower this
limit to $z \simlt 2.2$.  These limits are further supported by the
relatively strong continuum detections of the OT in the $B$ band (see
Table \TableMagnitudes) and in the $U$ band\refto{gcn214}, which
suggest little or no absorption by the intergalactic gas.

\refis{gcn219}
        Hjorth, J., Andersen, M. I., Cairos, L. M., Caon, N., Zapatero
        Osorio, M., Pedersen, H., Lindgren, B., Castro Tirado, A. J.
        GRB 990123 Spectroscopic Redshifts.
        {\it G.C.N.} {\bf 219} (1999).

No other convincing absorption features (aside from the normal telluric
absorption) are detected in our spectrum in the wavelength range
$\lambda \sim 4700$ -- $9000$ \AA. There are no obvious strong emission
lines out to $\lambda \sim 9600$ \AA.  The absence of other absorption
systems in the OT spectrum is not surprising: in this redshift range
there is about one metallic line absorber per unit redshift
interval\refto{Steidel92}.

\refis{Steidel92}
	Steidel, C. C. \&\ Sargent, W. L. W.
	Mg II Absorption in the Spectra of 103 QSOs: Implications for 
	the Evolution of Gas in High-Redshift Galaxies.
	Astroph.~J.~Suppl.~Ser. {\bf 80}, 1--108 (1992).

Our slit also included the bright galaxy $\sim 10$ arcsecond to the
West of the OT.  We determine a redshift of  $z= 0.2783 \pm 0.0005$ for
this galaxy, in agreement with the NOT measurement\refto{gcn249}. From
our IR observations we note that the $K$ band magnitude of this galaxy
is  $K = 16.39 \pm 0.03$.  At this redshift, this corresponds to a
normal, $L \sim L_*$ galaxy.

\refis{gcn249}
        Hjorth, J., Andersen, M. I., Pedersen, H., Zapatero-Osorio,
        M. R., Perez, E., Castro Tirado, A. J.
        GRB 990123 NOT Spectrum Update.
        {\it G.C.N.} {\bf 249} (1999).

\bigskip
\noindent{\bf The energetics of the burst}
\medskip

The combination of a high fluence and a redshift, $z\geq 1.6$, imply
that this burst is extremely energetic. For this discussion, we assume
a standard Friedmann model cosmology with $H_0 = 65$ km s$^{-1}$
Mpc$^{-1}$, $\Omega_0 = 0.2$, and $\Lambda_0 = 0$ (if $\Lambda_0 > 0$
then the inferred distance would increase, further aggravating the
energetics of the burst).  At the redshift $z = 1.6004$, the derived
luminosity distance is $D_L = 3.7 \times 10^{28}$ cm, corresponding to
the distance modulus $(m-M) = 45.39$, and 1 arcsec in projection
corresponds to 8.6 proper kpc, or 22.5 comoving kpc.

 From the observed fluence\refto{gcn224} (energy $> 20$ keV) of $5.1
\times 10^{-4}$ erg cm$^{-2}$ and the inferred $D_L$, we estimate the
$\gamma$-ray energy release, assuming the emission was isotropic, to
be $3.4 \times 10^{54}$ erg $\approx 1.9 M_\odot c^2$, more than the
rest mass energy of a neutron star.  The prompt optical flash detected
by ROTSE\refto{iauc7100a} had a $V$ magnitude of 8.9, implying a peak
luminosity in the UV band (restframe) of $\sim 3.3 \times 10^{16}\,
L_\odot$.  This is about a million times the luminosity of a normal
galaxy, and about a thousand times the luminosity of the brightest
quasars known!  The prompt optical emission is attributed to emission
from the reverse shock\refto{MR97,SP99}; in contrast, the late-time
afterglow which is the focus of this paper arises from the forward
shock.

\refis{iauc7100a}
        Akerlof, C. W. \&\ MacKay, T. A.
        GRB 990123.
        {\it I.A.U.C.} {\bf 7100} (1999).

\refis{SP99}
	Sari, R. \&\  Piran, T.
	GRB 990123, The Optical Fash and The Fireball Model.\hfill\break
	http://xxx.lanl.gov, astro-ph/9902009 (1999).

The above energetics and the peak luminosities are truly staggering.
Popular models suggest that GRBs are associated with stellar deaths,
and not with quasars (or the nuclei of galaxies). Perhaps the strongest
observational evidence in favor of this presumption is that some GRBs
are found offset from their host galaxy, and are therefore not
positionally consistent with an active nucleus. An isotropic energy
release of $1.9M_\odot c^2$ is, however, essentially incompatible with
the popular stellar death models (coalescence of neutron stars and the
currently popular model of the death of massive stars). This 
large energy release is barely
consistent with exotic models such as that of baryon
decay\refto{Pen98}; in such models essentially the entire rest mass
energy of the neutron star is released.

\refis{Pen98} Pen, U.-L. \&\ Loeb, A.,
	Gamma-ray Bursts from Baryon Decay in Neutron Stars.
	{\it Astrophys. J.} {\bf 509}, 537--543 (1998).

There are two ways to reduce the estimated energy release.  The
first possibility is that the burst is amplified by gravitational
lensing. Indeed, on the basis of early reports\refto{gcn206} of a
possible foreground galaxy near the line of sight to the OT, we
proposed that GRB 990123 was lensed. This hypothesis
received a boost when the NOT team reported\refto{gcn219} possible
intervening systems at redshifts of 0.286 and 0.201  which could be
potential candidates for lensing. Our own subsequent, deeper imaging
observations have failed to confirm any foreground galaxy and, as
discussed above, there is little spectroscopic evidence for the 0.286
and 0.201 absorption systems. There is therefore no direct evidence for
lensing at this time.

\refis{gcn206}
        Bloom, J. S., Gal, R. R., Lubin, L. L., Mulchaey, J. S.,
        Odewahn, S. C., \& Kulkarni S. R.
        GRB 990123 Optical Follow-up.
        {\it G.C.N.} {\bf 206} (1999).

Furthermore, the lensing probability is $\propto A^{-2}$ where $A$ is
the amplification provided by the lens. Even with $A\sim 2$, the expected
probability for lensing at a redshift of 1.6 is $10^{-3}$ (ref.
\Ref{Holz99}).  This probability is consistent with the observation
that roughly 1 in 500 cosmic radio sources, which likely have a similar
redshift distribution to GRBs, are lensed.  Thus it is not very
likely\refto{gcn241} that one out of the 15 bursts observed by BeppoSAX
so far would be highly magnified.

\refis{Holz99} Holz, D. E., Coleman, M. M. \&\ Quashnock, J. M. 
       Gravitational Lensing Limits on the Average Redshift of Gamma-ray
       Bursts.
       {\it Astrophys. J.} {\bf 510}, 54--63 (1999).

\refis{gcn241}
        Schaefer, B. E.
        GRB990123, Probability of gravitational lensing.
        {\it G.C.N.} {\bf 241} (1999).

We now consider the second possibility: the emitting surface in GRB
990123 is not spherical.  Indeed, almost all energetic sources in
astrophysics (e.g. quasars and accreting stellar black holes, pulsars)
are not spherically emitting sources but display jet-like geometry.
Not surprisingly there is an extensive literature on jets in the GRB
context (e.g. refs.  \Ref{MHIM93,MaoYi94,Katz97,MR97A,Dar98}) as well
as for the afterglow emission\refto{Rhoads97}.

\refis{MHIM93} Mochkovich, R., Hernanz, M., Isern, J. \&\ Martin, X.
               Gamma-ray bursts as collimated jets from neutron star/black
               hole mergers.
               {\it Nature} {\bf 361}, 236--237, (1993).

\refis{MaoYi94} Mao, S., \& Yi, I.
                Relativistic beaming and gamma-ray bursts.
                {\it Astrophys. J.} {\bf  424}, L131--134, (1994).

\refis{Katz97} Katz, J. I.
               Yet Another Model for Gamma Ray Bursts.
               {\it Astrophys. J.} {\bf  490}, 633--640, (1997).

\refis{MR97A}  M\'esz\'aros, P., \& Rees, M.J.
               Poynting Jets from Black Holes and Cosmological Gamma-Ray
               Bursts.
               {\it Astrophys. J.} {\bf 482}, L29--32, (1997).

\refis{Dar98} Dar, A.
	Can Fireball Models Explain Gamma-ray Bursts?
	{\it Astrophys. J.} {\bf 500}, L93-L96 (1998).

\refis{Rhoads97} Rhoads, J.E.
               How to Tell a Jet from a Balloon: A Proposed Test for
	       Beaming in Gamma-Ray Bursts.
               {\it Astrophys. J.} {\bf 478}, L1--L4, (1997).

Should the emitting surface be a jet with an opening angle of radius
$\theta_0$ then the inferred energy is reduced from the isotropic
value by the beaming factor, $f_b \sim \theta_0^2/2$. Here we make the
reasonable assumption based on other astrophysical sources that the
jet is two sided.  Thus if $\theta_0\sim 0.3$ then $f_b\sim 5\times
10^{-2}$, and the $\gamma$-ray energy released is $\sim 10^{53}$ erg,
a value within reach of current models for the origin of GRBs.

As discussed below, a marked steepening of the afterglow emission is
the clearest and simplest signature of a jet-like geometry.  The three
well studied GRBs
(970228, 970508, 980703)\refto{Sokolov98,Pedersen98,Galama98,Bloom98,
Fruchter98,Frail99} exhibit, at late
times (epochs longer than few days) a single power law decay and with
indicies that are reasonable for spherical expansion, $\alpha
\sim -1.2$.  (The complicated early time variations seen in afterglows
of many GRBs are not relevant to the discussion here.) The radio
afterglow of GRB 970508 has been used to argue\refto{WKF98} for
beaming in this source but this result is model dependent.  It is
against this backdrop that we now proceed to see if there is evidence
for non-spherical geometry in GRB 990123.

\refis{Pedersen98}
	Pedersen, H. et al.
	Evidence for Diverse Optical Emission from Gamma-Ray Burst
        Sources.
	{\it Astrophys. J.} {\bf 496}, 311--315 (1998).
\refis{Galama98}
	Galama, T. et al. 
	Optical Follow up of GRB 970508.
	{\it Astrophys. J.} {\bf 497}, L13--L16 (1998).
\refis{Bloom98}
	Bloom, J. S., Djorgovski, S. G., Kulkarni, S. R. \&\
	Frail, D. A. 
	The Host Galaxy of GRB 970508.
	{\it Astrophys. J.} {\bf 507}, L25--L28 (1998). 	
\refis{Fruchter98}
	Fruchter, A. S. et al.
	The Fading Optical Counterpart of GRB 970228, Six Months
	and One Year Later.
	http://xxx.lanl.gov, astro-ph/9807295 (1998).
\refis{Frail99}
	Frail, D. A., Bloom, J. S., Kulkarni, S. R. \&\ Taylor, G. B. 
	The Light Curve and the Spectrum of the Radio Afterglow of GRB 980703.
	In preparation, (1999).	
\refis{WKF98} 
	Waxman, E., Kulkarni, S. R. \&\ Frail, D. A. 
       Implications of the Radio Afterglow from the Gamma-Ray Burst of 
       1997 May 8.
       {\it Astrophys. J.} {\bf 497}, 288 (1998).

\bigskip
\noindent{\bf The shape of the burst: beaming?}
\medskip

Our knowledge of the shape of the emitting region in GRBs is limited
because, due to relativistic beaming, only a small portion (angular
size $\sim \Gamma^{-1}$, where $\Gamma$ is the Lorentz factor of the
bulk motion of the emitting material) is visible to the observer. Thus
the observer is unable to distinguish a sphere from a jet as long as
$\Gamma > \theta_0^{-1}$.  However, as the source continues its radial
expansion, $\Gamma$ will decrease, and when $\Gamma < \theta_0^{-1}$
there will be a marked decrease in the observed flux.  Even if the jet
continues to evolve as a cone with a constant opening angle, the
observer will see the flux reduced by the ratio of the solid angle of
the emitting surface ($\propto \theta_0^2$) to that expected for a
spherical expansion ($\propto \gamma^{-2}$). Thus the light curve
will steepen by $\gamma^2\theta_0^2 \propto t^{-3/4}$. This 
is a purely geometric effect. 

Rhoads\refto{Rhoads99} makes the important point that unless the jet
is confined by some mechanism (as advocated in ref. \Ref{Dar98}) the
jet will also expand sideways at the sound speed of shocked
relativistic material, $c_s=c/\sqrt{3}$.  Thus as the jet evolves, the
solid angle of the emitting region increases as
$\pi(\theta_0+c_st_{\rm co}/ct)^2$, where $t$ is the time since the
burst in the GRB frame, and $t_{\rm co}=t/\Gamma$ is the elapsed time
in the frame moving with the jet.  When $\Gamma$ falls below
$\theta_0^{-1}$ the jet spreads in the lateral direction.
Consequently, in this regime, $\Gamma$ decreases exponentially with
radius $r$ instead of as a power law.  We can therefore take $r$ to be
essentially constant during this spreading phase. Since $t_E \approx
r/2 \Gamma^2 c$, we find $\Gamma \propto t_E^{-1/2}$ during the
spreading phase; here, $t_E$ is the elapsed time in the frame of the
observer at Earth. With this scaling the typical synchrotron
frequency, $\nu_m$, decreases as $t_E^{-2}$ and the flux at this
frequency drops as $t_E^{-1}$. The flux at a fixed frequency above
$\nu_m$ drops as $F_{\nu}=F_{\nu,m}(\nu/\nu_m)^{-(p-1)/2} \propto
t_E^{-p}$, where $p\sim 2.5$ is the electron energy power-law index.
This is more than one power of $t$ steeper than that for a spherical
afterglow, $t_E^{-3(p-1)/4} \propto t_E ^ {-1.1}$.

\refis{Rhoads99} Rhoads, J.E.
               Constraining Gamma Ray Burst Beaming.
               {\it Preprint}, (1998).

In contrast to this discussion, we expect another kind of break and
that is when the electrons responsible for the photons in the band of
interest enter the cooling regime. In this case, $\alpha$ decreases by
$1/4$ (e.g.~ref.  \Ref{Sari98}). However, unlike the break due to jet
geometry (discussed above) this break is not broad-band. Regardless,
in both cases, electron cooling or jets, $\alpha$ is expected to evolve over a
timescale of $t_{\rm break}$. Thus the expected change in $\alpha$,
$\Delta\alpha\equiv\alpha_1-\alpha_2$  is $3/4$ (constant $\theta_0$),
$1-\alpha_1/3$ (spreading jet) or  $1/4$ (electron cooling) should
be considered to be  {\it
upper limits;} see also ref. \Ref{MR99}.

\refis{MR99}
	M\'esz\'aros, P. \&\ Rees, M. J. 
	GRB 990123: Reverse and Internal Shock Flashes and Late
	Afterglow Behavior.
	http://xxx.lanl.gov, astro-ph/9902367 (1999).

As can be seen from Figure \Light-Curves we do indeed see a break in
the $r$-band flux with $\Delta\alpha_r= 0.55\pm 0.07$; the 95\%
confidence interval, taking into account of covariance with
the epoch of the break, is 0.4--0.7 (see Figure \Light-Curves).
On statistical grounds, we can rule out
$\Delta\alpha=1/4$.  The prediction of 
$\Delta\alpha=1/4$ for cooling
electron cooling arises from basic synchrotron theory and is therefore
a robust prediction. Therefore, we firmly conclude that 
the observed break is not due to electron cooling.

The measured value of $\Delta\alpha$ support the hypothesis of a jet in
GRB 990123. However, the lack of a break in the $K$-band light curve is
an issue of considerable concern.  As explained in the legend to Figure
\Light-Curves, a single power-law model provides a good fit to the
$K$-band data.  Within statistical errors we can state that the
$r$-band model does not conform to the $K$-band light curve.
Unfortunately, the quality and quantity of the $K$-band data and the
uncertainty in the adopted flux of the host galaxy do not permit us to
derive parameters for a broken power law model independent of those
derived from the $r$-band light curve.  [Future $K$-band observations
when the OT has essentially disappeared would help us resolve at least
the latter uncertainty.]

We cautiously advance the hypothesis that we are seeing evidence for a
jet (perhaps with no sideways expansion). If so, $\theta_0\sim
\Gamma(t_{\rm break})^{-1}$ and for typical parameters $\theta_0\sim
0.2$ in which case $f_b\sim 0.02$. Thus the energy released in
$\gamma$-rays alone is $6\times 10^{52}$ erg.  Including the X-ray
afterglow\refto{Heise99} raises the total energy release to $10^{53}$
erg. GRB 990123 was indeed a very energetic GRB.  Following the
discovery of the afterglow phenomenon\refto{Costa97, jvP97,Frail97}, it
was thought that afterglow observations would provide only global
parameters of the GRB explosion (e.g. energy released, ambient
density), but it now appears that afterglow observations may give us
insight into the geometry of the explosion as well.



\bigskip
\bigskip

\noindent{\bf Acknowledgment.} 
We are indebted to G. Neugebauer for obtaining IR data on 29 January 1999.
We thank A. Filippenko, L. Hillenbrand  and J. Carpenter for kindly
agreeing to exchange telescope time thereby enabling us to follow up
this very interesting GRB and B. Oppenheimer for help with
observations on February 7 and 8.
Some of the observations reported here were obtained at the W. M. Keck
Observatory, which is operated by the California Association for
Research in Astronomy, a scientific partnership among California
Institute of Technology, the University of California and the National
Aeronautics and Space Administration.  It was made possible by the
generous financial support of the W. M. Keck Foundation.  
SRK's research is supported by the National Science
Foundation and NASA.  SGD acknowledges a partial support from the
Bressler Foundation.  

\vfill\eject



\centerline{\bf Table \TableMagnitudes.
                Photometric observations of GRB 990123}
\smallskip
\hrule
\smallskip
\settabs\+ 1234567890123 & 1234567890 & 12345 & 12345678901234 &
12345 &.\cr

\+ Date
& Telescope
& Band
& ~~Magnitude
& Reference/
& \cr

\+  (UT)
&
&
&
& Observers
&
& \cr

\smallskip
\hrule
\medskip

\+ Jan 23.577   & P60  & ~r &  $18.65 \pm 0.04$  & SCO & \cr
\+ Jan 23.578   & P200 & ~B &  $18.93 \pm 0.04$  & LML, JSM & \cr
\+ Jan 23.958   & UPSO & ~r &  $20.00 \pm 0.05$  & ref. \Ref{Sagar99} & \cr
\+ Jan 24.005   & UPSO & ~B &  $20.16 \pm 0.15$  & ref. \Ref{Sagar99} & \cr
\+ Jan 24.18    & BAO  & ~r &  $20.39 \pm 0.15^{a}$ & ref. \Ref{gcn233} & \cr
\+ Jan 24.194   & BAO  & ~B &  $20.64 \pm 0.07$     & ref. \Ref{gcn233} & \cr
\+ Jan 24.547   & WO   & ~r &  $20.93 \pm 0.2^{a}$ & ref. \Ref{gcn215} & \cr
\+ Jan 24.636   & KI   & ~K &  $18.29 \pm 0.04$     & Koerner \&\
Kirkpatrick & \cr
\+ Jan 24.934   & UPSO & ~r &  $21.22 \pm 0.08^{a}$ &  ref. \Ref{Sagar99}& \cr
\+ Jan 24.978   & UPSO & ~B &  $22.06 \pm 0.20$     & ref.  \Ref{Sagar99}& \cr
\+ Jan 25.14    & BAO  & ~r &  $21.37 \pm 0.15^{a}$ & ref. \Ref{gcn233} & \cr
\+ Jan 25.940   & UPSO & ~r &  $21.73 \pm 0.12^{a}$ & ref. \Ref{Sagar99} & \cr
\+ Jan 26.154   & BAO  & ~r &  $22.09 \pm 0.10^{a}$ & ref. \Ref{gcn233} & \cr
\+ Jan 27.652   & KI   & ~K &  $19.73 \pm 0.07$  &  MAM, ISM, HIT & \cr
\+ Jan 29.677   & KI   & ~K &  $20.40 \pm 0.08$  &  G. Neugebauer \&\ LA & \cr
\+ Jan 30.52    & MDM  & ~r &  $23.28 \pm 0.18^{a}$ & JH, IAY  & \cr
\+ Feb 3.54     & MDM  & ~r &  $23.88 \pm 0.24^{a}$ & JH, IAY  & \cr
\+ Feb 6.6      & KI   & ~K &  $20.84 \pm 0.13$  & NK  & \cr
\+ Feb 7.610    & KI   & ~K &  $21.00 \pm 0.09$  & SRK & \cr
\+ Feb 8.6      & KI   & ~K &  $20.83 \pm 0.11$  & SRK & \cr
\+ Feb 9.052    & HST  & ~r &  $24.12 \pm 0.10$  & ref. \Ref{Bloom99} & \cr
\+ Feb 9.6      & KI   & ~K &  $20.97 \pm 0.10$  & L. Hillenbrand & \cr
\+ Feb 10.6     & KI   & ~K &  $21.21 \pm 0.11$   & L. Hillenbrand & \cr
\+ Feb 9.654    & KII  & ~r &  $23.91 \pm 0.07^{a}$ & A. Filippenko & \cr
\+ Feb 14.50    & MDM  & ~r &  $24.10 \pm 0.10^{a}$ & JH, IAY & \cr

\medskip
\hrule
\smallskip

\medskip
\noindent Notes:

\smallskip
\noindent (1) Telescope Acronyms. 
P60, Palomar 60-inch telescope.
P200, Palomar 200-inch Hale telescope.
UPSO, U.~P.~State Observatory 104-cm telescope.
BAO, Bologna Astronomical Observatory 1.5-m telescope.
WO, Fred L.~Whipple Observatory 1.2-m telescope.
KI, Keck-I 10-m telescope.
MDM, 2.4-m Hiltner Telescope.
KII, Keck-II 10-m telescope.
HST, Hubble Space Telescope.

\refis{Sagar99}
        Sagar, R., Pandey, A. K., Mohan, V., Yadav, R. K. S.,
        Nilakshi, Bhattacharya, D. \&\ Castro-Tirado, A. J.
        Optical follow up of the GRB 990123 source from UPSO,
        Nainital.
        {http://xxx.lanl.gov}, astro-ph/9902196 (1999).

\refis{gcn233}
        Masetti, N., Palazzi, E., Pian, E., Frontera, F., Bartolini,
        C., Guarnieri, A., Piccioni, A., \& Costa, E.
        GRB990123, Optical BVRI Observations.
        {\it G.C.N.} {\bf 233} (1999).

\refis{gcn215}
        Garnavich, P., Jha, S., Stanek, K., Garcia, M.
        GRB990123, optical observation.
        {\it G.C.N.} {\bf 215} (1999).

\smallskip \noindent (2) Photometric zeropoint.  The night of 23
        January UT was photometric at Palomar.  Absolute zero-points
for the Gunn-$r$ and $B$-band data were obtained using the
observations of standard star-fields\refto{Thuan76,Oke83,Landolt92}.
The absolute zero-point for the $K$-band observations was obtained
using Keck-I data from the night of 24 January UT. Each bandpass
zero-point was confirmed independently by at least two members of our
group. All subsequent imaging by our group were reduced in the
standard manner and photometry was propagated using a set of secondary
stars in the GRB field.  Photometry derived from data taken in the
Cousins $R$-band were recalibrated to Gunn-$r$; the quoted
uncertainties include the estimated zeropoint and statistical
uncertainties.  In this table, we include all of our photometric data
obtained to date and also data drawn from the literature which we
could reliably place in our $B$, Gunn-$r$, and $K$-band system.
Magnitudes marked with the superscript $a$ are $R_c$ magnitudes
corrected to Gunn r-band using secondary reference stars reported in
ref.~\Ref{gcn204,gcn206}.

\refis{Thuan76}
        Thuan, T. X., \& Gunn, J. E.
        A new four-color intermediate-band photometric system.
        {\it Proc.~Astron.~Soc.~Pac.} {\bf 88}, 543--547 (1976).

\refis{Landolt92}
        Landolt, A.
        UBVRI photometric standard stars in the magnitude range
        11.5--16.0 around the celestial equator.
        {\it Astron.~J.} {\bf 104}, 340--376 (1992).

\refis{Oke83}
        Oke, J. B., \& Gunn, J. E.
        Secondary standard stars for absolute spectrophotometry.
        {\it Astrophys. J.} {\bf 266}, {713--717} (1983).

\refis{gcn204}
        Zhu, J., \& Zhang, H. T.
        GRB990123 Optical Observation
        {\it G.C.N.} {\bf 204} (1999).

\vfill \eject

\centerline{Table \Table-Lines. 
            Absorption Lines Detected in the Spectrum of the OT}
\bigskip

\smallskip
\hrule
\smallskip

\settabs\+ 12345678901234 & 1234567890 & 1234567890 & 123456789 & 123456 &
12345 &.\cr

\+ Line ID
& $\lambda_{obs,air}$
& $\lambda_{rest,vac}$
& ~~~$z$
& $W_{\lambda,obs}$
& ~$\pm$
& \cr

\+
& ~~(\AA)
& ~~(\AA)
&
& ~(\AA)
& ~(\AA)
& \cr

\smallskip
\hrule
\medskip

\+ Al III 1862 &  4843.74 &  1862.78  & 1.6010 &  1.23 & 0.07 & \cr
\+ Zn II 2026  &  5267.29 &  2026.14  & 1.6004 &  2.08 & 0.17 & \cr
\+ Cr II 2062  &  5361.77 &  2062.23  & 1.6007 &  1.26 & 0.10 & \cr
\+ Zn II 2062  &  5361.77 &  2062.66  & 1.6002 &  1.26 & 0.10 & \cr
\+ Fe II 2260  &  5877.17 &  2260.78  & 1.6003 &  0.91 & 0.07 & \cr
\+ Fe II 2344  &  6096.14 &  2344.21  & 1.6012 &  3.04 & 0.28 & \cr
\+ Fe II 2374  &  6173.87 &  2373.73  & 1.6016 &  3.07 & 0.28 & \cr
\+ Fe II 2382  &  6195.29 &  2382.76  & 1.6008 &  3.60 & 0.40 & \cr
\+ Fe II 2586  &  6725.75 &  2586.64  & 1.6009 &  2.84 & 0.37 & \cr
\+ Fe II 2600  &  6759.94 &  2600.18  & 1.6005 &  3.48 & 0.14 & \cr
\+ Mg II 2796  &  7269.47 &  2796.35  & 1.6003 &  4.59 & 0.30 & \cr
\+ Mg II 2803  &  7289.49 &  2803.53  & 1.6008 &  4.80 & 0.52 & \cr
\+ Mg I 2852   &  7416.97 &  2852.97  & 1.6005 &  2.87 & 0.18 & \cr

\medskip
\hrule
\smallskip

\medskip
\noindent Notes:

\smallskip
\noindent (1)
$W_{\lambda,obs}$ is the observed line equivalent width; for the restframe
values, divide by $(1+z)$.

\smallskip
\noindent (2)
The observed absorption line at 5362 \AA\ is a blend of two lines,
Cr II 2062 and Zn II 2062, and thus it appears twice.

\vfill\eject


\def\arcmin{\hbox{$^\prime$}}
\def\arcsec{\hbox{$^{\prime\prime}$}}
\def\fd{\hbox{$~\!\!^{\rm d}$}}
\def\fh{\hbox{$~\!\!^{\rm h}$}}
\def\fm{\hbox{$~\!\!^{\rm m}$}}
\def\fs{\hbox{$~\!\!^{\rm s}$}}

\noindent{\bf Figure \Discovery-Image.} Discovery image
of the optical counterpart of GRB 990123. (Left) A 4 arcmin $\times$ 4
arcmin portion centered on the OT and extracted from the Digital
Palomar Observatory Sky Survey II (DPOSS); the DPOSS is the digitized
version of the second Palomar Observatory Sky Survey.  (Right) A CCD
image in the Gunn-$r$ band obtained at the Palomar 60-inch.  The new
source, the presumed optical afterglow of GRB 990123, is marked.  The
circle is the 50-arcsecond localization of the X-ray
afterglow\refto{Heise99} as obtained from the Narrow Field Instrument
aboard BeppoSAX.  Absolute astrometry on the Palomar 60-inch discovery
image was obtained by comparison of 34 objects near the optical
transient with the USNO-A2.0 V2.0 Catalogue\refto{Monet98}.  The
r.m.s.~uncertainties of the astrometry are 0.28\arcsec (R.A.) and
0.26\arcsec (declination).  We find the position of the optical
transient to be, \hbox{$\alpha = $ 15\fh 25\fm 30\fs .34},
\hbox{$\delta$ = +44\fd 45\arcmin 59\arcsec .11} (J2000).

\refis{Monet98}
        Monet, D. G.
        The 526,280,881 Objects In The USNO-A2.0 Catalog.
        {\it Bull.~Amer.~Astron.~Soc.} {\bf 193}, \hbox{\#120.03} (1998).

\vfill\eject


\noindent{\bf Figure  \Light-Curves.}
Optical and infrared light curves of the transient afterglow of GRB 990123
from 4 hours to 23 days after the burst.  The observed magnitudes from
Table 1 have been corrected for Galactic extinction and converted to flux
(see below).  

\noindent ({\it Top})
The light curve of the transient + host galaxy.  From
ref.~\Ref{Schlegel98} we estimate the Galactic extinction in the
direction of the optical transient $(l,b = 73.12^\circ, 54.64^\circ)$
to be E(B$-$V) = 0.01597. Thus, assuming the average Galactic
extinction curve ($R_V$ = 3.1), the extinction measure is $A_B =
0.069$, $A_r$ = 0.041, and $A_K = 0.006$ mag.

\refis{Schlegel98}
        Schlegel, D. J., Finkbeiner, D. P., \& Davis, M.
        Maps of Dust Infrared Emission for Use in Estimation of
        Reddening and Cosmic Microwave Background Radiation Foregrounds.
        {\it Astrophys.~J.} {\bf 500}, 525--553 (1998). 

\noindent ({\it Bottom})
The inferred light curve of the transient. The contribution of the
host flux has been subtracted from the Gunn-$r$ and $K$-band fluxes at
each epoch. The adopted host galaxy fluxes are $K=22.1\pm0.3$
magnitude corresponding to $0.9 \pm 0.3\, \mu$Jy and $0.58 \pm
0.04\, \mu$Jy (Gunn r); see below for details. The error bars include
photometric error as well as the uncertainties in the estimated host
galaxy flux.

{\it $r$-band curve.} A single power law decay model is inconsistent
with the $r$-band data ($\chi^2_{\rm min}/d.o.f \simeq 10$).  We then
fit the $r$-band points to a broken power law model: $F_\nu = F_*
(t/t_{\rm break})^{\alpha_{1r}}$ for $t \leq t_{\rm break}$ and $F_\nu
= F_* (t/t_{\rm break})^{\alpha_{2r}}$ for $t \geq t_{\rm break}$;
here, $t$ is the time in days since the GRB and $t_{\rm break}$ is
epoch of the break.  Using a Levenberg-Marquardt $\chi^2$
minimization\refto{Press92}, we find the $r$-band data are adequately
fit by this broken power-law model ($\chi^2 = 12.1$ for 9 d.o.f.) with
the following parameters $F_* = 7.0 \pm 1.9 \,\mu$Jy, $t_{\rm break}
= 2.04 \pm 0.46$ d, $\alpha_{1r} = -1.10 \pm 0.03$, and $\alpha_{2r} =
-1.65 \pm 0.06$.  The errors quoted are 1-$\sigma$ confidence
intervals for each respective parameter.  Our value of $\alpha_{1r}$
agrees very well with the Sagar et
al.'s\refto{Sagar99} determination ($\alpha =-1.10\pm 0.06$),
based on $B$ and $R$ observations obtained over the
first two nights.
There is some covariance
between $\alpha_2$ and $t_{\rm break}$ so that the confidence level on
a subset of parameters may extend beyond the above quoted errors.  For
instance, concurrent values of ($t_{\rm break}$ = 1 d, $\alpha_{2r} =
-1.5$) are allowed at the 2-$\sigma$ level as are ($t_{\rm break}$ = 3
d, $\alpha_{2r} = -1.8$).

\refis{Press92}
	Press, W. H., Teukolsky, S. A., Vetterling, W. T.
	\&\ Flannery, B. P. 
	{\it Numerical Recipes in C.} pp 683--699.
	Cambridge University Press, New York (1992).

{\it $K$-band curve.} In contrast, the $K$-band data can be modelled
with a single power law index: $\alpha_K=-1.12 \pm 0.11$ ($\chi^2=8.8$
for 6 d.o.f.). This is an acceptable fit (probability of 35\%).  The
$K$-band data cannot be statistically reconciled to the best fit
$r$-band model (i.e. assuming $\alpha_{1r}$, $\alpha_{2r}$ and $t_{\rm
break}$ to be fixed) with $\chi^2$ ranging from 28 to 66 (7 d.o.f), depending
on the value assumed for the flux of the host galaxy.

The flux of the host galaxy was estimated as follows.  For the $r$-band
observations, we used the HST/STIS imaging where the host galaxy is
clearly resolved and thus the flux of the OT and the host can be
measured quite accurately.  In ref. \Ref{Bloom99} we discuss in great
detail the tie between Gunn-$r$ photometry and the STIS photometry
(done with CLEAR filter which corresponds approximately to the V band)
as well as $R$-band photometry. The photometric ties are robust so that
the flux of the host in $r$ band is well determined, $0.58\pm
0.04\,\mu$Jy. 

We determined the flux of the host galaxy in K-band as follows.  We
used the deepest exposures with the best seeing images (February 9 and
February 10). In these images the OT is clearly resolved from the host
galaxy.  Sets of pixels dominated by the OT or by the galaxy were
masked, and total fluxes with such censored data were evaluated in
photometric apertures of varying radii.  Total fluxes of the OT+galaxy
were also measured in the same apertures using the uncensored data.
We also varied the aperture radii, and the position and the size of
the sky measurement annulus.  On February 9 (February 10) UT we find
the OT contributes 65 (57) percent ($\pm 10$ percent) of the total
OT+galaxy light.  The estimated errors of the fractional contributions
of the OT to the total light reflect the scatter obtained from
variations in the parameters of these image decompositions.  In both
epochs the fractional contribution of the host implies a flux of the
host galaxy is $0.9 \pm 0.3\, \mu$Jy ($K_{\rm host} = 22.1 \pm 0.3$
mag).  As a further test, we fit the K-band light curve assuming the
flux is given as the sum of a power-law (OT) plus a constant flux
(host galaxy).  This fit  yields $K_{\rm host} = 21.55\pm 0.20$
magnitude,
in agreement with the magnitude derived from direct imaging.
However, we note that the fit is barely acceptable ($\chi^2=11.9$,
5 d.o.f.).

\vfill\eject

\noindent{\bf Figure \Spectrum-Figure.}
The spectrum of the optical transient of GRB 990123.   Prominent
absorption features are marked; all of these can be attributed to
various absorption lines at a redshift, $z_{abs}=1.6004$; the two
features labeled as ``atm'' are telluric absorption features (A and B
bands) of the Earth's atmosphere.  

Three 600-s integrations were taken, though the last one was
terminated after 563 s due to the onset of fog; the integrations
commenced at UT 15:38, UT 15:49 and UT 16:01, respectively.  The last
spectrum also suffered from a higher sky background from the
brightening dawn sky.  The sky conditions appeared to be good during
the first two integrations, with seeing about 1 arcsecond.  We used
the 300 line/mm grating and the resulting wavelength coverage was 4700
\AA\ to 9000 \AA\ with an effective resolution of about 11.6\AA, (full width at half
maximum).  Spectra were taken with the 1.0 arcsecond wide longslit
oriented at position angle (PA) $90^\circ$ (i.e.~East-West) and
centered on the OT; the slit also included the prominent galaxy that
is about 10 arcsecond West of the OT (see Figure \Discovery-Image).

The usual calibration observations were not obtained given the urgency
of the observations.  The calibration of the wavelength was obtained
using a dispersion curve measured with the same grating and slit, but
a somewhat different tilt, ten days earlier.  Wavelengths of the
unblended night sky emission lines were then used to adjust the
zero-point of the wavelength solution fit; this procedure also
compensated flexure of the instrument between the exposures.  An
approximate flux calibration was accomplished by using the instrument
response curve also measured 10 days earlier; the instrument response
is stable enough for this approach.  The slit losses due to the seeing
and the difference of the actual slit PA from the optimal
(parallactic) angle at the time of observations make the net flux
calibration and the overall the shape of the spectrum somewhat
uncertain. Nonetheless, we note that our spectroscopic flux
measurement is in excellent agreement with photometry from
interpolation of the $r$-band light curve.  In any case, this is not
important for the redshift measurements discussed in the text.

\vfill\eject

\centerline {\bf References}
\bigskip

\endreferences

\vfill\eject

\endmode
\bye

%% file: reforder.tex
\catcode`@=11
\newcount\r@fcount \r@fcount=0
\newcount\r@fcurr
\immediate\newwrite\reffile
\newif\ifr@ffile\r@ffilefalse
\def\w@rnwrite#1{\ifr@ffile\immediate\write\reffile{#1}\fi\message{#1}}

\def\writer@f#1>>{}
\def\referencefile{
  \r@ffiletrue\immediate\openout\reffile=\jobname.ref%
  \def\writer@f##1>>{\ifr@ffile\immediate\write\reffile%
    {\noexpand\refis{##1} = \csname r@fnum##1\endcsname = %
     \expandafter\expandafter\expandafter\strip@t\expandafter%
     \meaning\csname r@ftext\csname r@fnum##1\endcsname\endcsname}\fi}%
  \def\strip@t##1>>{}}

\def\citeall#1{\xdef#1##1{#1{\noexpand\cite{##1}}}}
\def\cite#1{\each@rg\citer@nge{#1}}	

\def\each@rg#1#2{{\let\thecsname=#1\expandafter\first@rg#2,\end,}}
\def\first@rg#1,{\thecsname{#1}\apply@rg}	
\def\apply@rg#1,{\ifx\end#1\let\next=\relax
\else,\thecsname{#1}\let\next=\apply@rg\fi\next}

\def\citer@nge#1{\citedor@nge#1-\end-}	
\def\citer@ngeat#1\end-{#1}
\def\citedor@nge#1-#2-{\ifx\end#2\r@featspace#1 
  \else\citel@@p{#1}{#2}\citer@ngeat\fi}	
\def\citel@@p#1#2{\ifnum#1>#2{\errmessage{Reference range #1-#2\space is bad.}%
    \errhelp{If you cite a series of references by the notation M-N, then M and
    N must be integers, and N must be greater than or equal to M.}}\else%
 {\count0=#1\count1=#2\advance\count1 by1\relax\expandafter\r@fcite\the\count0,%
  \loop\advance\count0 by1\relax
    \ifnum\count0<\count1,\expandafter\r@fcite\the\count0,%
  \repeat}\fi}

\def\r@featspace#1#2 {\r@fcite#1#2,}	
\def\r@fcite#1,{\ifuncit@d{#1}
    \newr@f{#1}%
    \expandafter\gdef\csname r@ftext\number\r@fcount\endcsname%
                     {\message{Reference #1 to be supplied.}%
                      \writer@f#1>>#1 to be supplied.\par}%
 \fi%
 \csname r@fnum#1\endcsname}
\def\ifuncit@d#1{\expandafter\ifx\csname r@fnum#1\endcsname\relax}%
\def\newr@f#1{\global\advance\r@fcount by1%
    \expandafter\xdef\csname r@fnum#1\endcsname{\number\r@fcount}}

\let\r@fis=\refis			
\def\refis#1#2#3\par{\ifuncit@d{#1}
   \newr@f{#1}%
   \w@rnwrite{Reference #1=\number\r@fcount\space is not cited up to now.}\fi%
  \expandafter\gdef\csname r@ftext\csname r@fnum#1\endcsname\endcsname%
  {\writer@f#1>>#2#3\par}}

\def\ignoreuncited{
   \def\refis##1##2##3\par{\ifuncit@d{##1}%
     \else\expandafter\gdef\csname r@ftext\csname r@fnum##1\endcsname\endcsname%
     {\writer@f##1>>##2##3\par}\fi}}

\def\r@ferr{\endreferences\errmessage{I was expecting to see
\noexpand\endreferences before now;  I have inserted it here.}}
\let\r@ferences=\references
\def\references{\r@ferences\def\endmode{\r@ferr\par\endgroup}}

\let\endr@ferences=\endreferences
\def\endreferences{\r@fcurr=0
  {\loop\ifnum\r@fcurr<\r@fcount
    \advance\r@fcurr by 1\relax\expandafter\r@fis\expandafter{\number\r@fcurr}%
    \csname r@ftext\number\r@fcurr\endcsname%
  \repeat}\gdef\r@ferr{}\endr@ferences}


\let\r@fend=\endpaper\gdef\endpaper{\ifr@ffile
\immediate\write16{Cross References written on []\jobname.REF.}\fi\r@fend}

\catcode`@=12

\citeall\refto		
\citeall\ref		%
\citeall\Ref		%

%% file: citmac.tex
\def\singlespace{\baselineskip 12pt \lineskip 1pt \parskip 2pt plus 1 pt}

\def\today{\number\day\enspace
     \ifcase\month\or January\or Febuary\or March\or April\or May\or
     June\or July\or August\or September\or October\or
     November\or December\fi \enspace\number\year}
\def\clock{\count0=\time \divide\count0 by 60
    \count1=\count0 \multiply\count1 by -60 \advance\count1 by \time
    \number\count0:\ifnum\count1<10{0\number\count1}\else\number\count1\fi}
\footline={\hss -- \folio\ -- \hss}

\def\deg{\ifmmode^\circ\else$^\circ$\fi}
\def\solar{\ifmmode_{\mathord\odot}\else$_{\mathord\odot}$\fi}
\def\jref#1 #2 #3 #4 {{\par\noindent \hangindent=3em \hangafter=1 
      \advance \rightskip by 5em #1, {\it#2}, {\bf#3}, #4.\par}}
\def\ref#1{{\par\noindent \hangindent=3em \hangafter=1 
      \advance \rightskip by 5em #1.\par}}
\newcount\eqnum
\def\nexteq{\global\advance\eqnum by1 \eqno(\number\eqnum)}
\def\lasteq#1{\if)#1[\number\eqnum]\else(\number\eqnum)\fi#1}
\def\preveq#1#2{{\advance\eqnum by-#1
    \if)#2[\number\eqnum]\else(\number\eqnum)\fi}#2}

\def\tableheight{\vrule width 0pt height 8.5pt depth 3.5pt}
{\catcode`|=\active \catcode`&=\active 
    \gdef\tabledelim{\catcode`|=\active \let|=\vbar
                     \catcode`&=\active \let&=\nobar} }
\def\table{\begingroup
    \def\twidth{\hsize}
    \def\tablewidth##1{\def\twidth{##1}}
    \def\defaultheight{\vrule width 0pt height 8.5pt depth 3.5pt}
    \def\heightdepth##1{\dimen0=##1
        \ifdim\dimen0>5pt 
            \divide\dimen0 by 2 \advance\dimen0 by 2.5pt
            \dimen1=\dimen0 \advance\dimen1 by -5pt
            \vrule width 0pt height \the\dimen0  depth \the\dimen1
        \else  \divide\dimen0 by 2
            \vrule width 0pt height \the\dimen0  depth \the\dimen0 \fi}
    \def\spacing##1{\def\defaultheight{\heightdepth{##1}}}
    \def\nextheight##1{\noalign{\gdef\tableheight{\heightdepth{##1}}}}
    \def\end{\cr\noalign{\gdef\tableheight{\defaultheight}}}
    \def\zerowidth##1{\omit\hidewidth ##1 \hidewidth}    
    \def\hline{\noalign{\hrule}}
    \def\skip##1{\noalign{\vskip##1}}
    \def\bskip##1{\noalign{\hbox to \twidth{\vrule height##1 depth 0pt \hfil
        \vrule height##1 depth 0pt}}}
    \def\header##1{\noalign{\hbox to \twidth{\hfil ##1 \unskip\hfil}}}
    \def\bheader##1{\noalign{\hbox to \twidth{\vrule\hfil ##1 
        \unskip\hfil\vrule}}}
    \def\spanloop{\span\omit \advance\mscount by -1}
    \def\extend##1##2{\omit
        \mscount=##1 \multiply\mscount by 2 \advance\mscount by -1
        \loop\ifnum\mscount>1 \spanloop\repeat \ \hfil ##2 \unskip\hfil}
    \def\vbar{&\vrule&}
    \def\nobar{&&}
    \def\hdash##1{ \noalign{ \relax \gdef\tableheight{\heightdepth{0pt}}
        \toks0={} \count0=1 \count1=0 \putout##1\end 
        \toks0=\expandafter{\the\toks0 &\end} \xdef\piggy{\the\toks0} }
        \piggy}
    \let\e=\expandafter
    \def\putspace{\ifnum\count0>1 \advance\count0 by -1
        \toks0=\e\e\e{\the\e\toks0\e&\e\multispan\e{\the\count0}\hfill} 
        \fi \count0=0 }
    \def\putrule{\ifnum\count1>0 \advance\count1 by 1
        \toks0=\e\e\e{\the\e\toks0\e&\e\multispan\e{\the\count1}\leaders\hrule\hfill}
        \fi \count1=0 }
    \def\putout##1{\ifx##1\end \putspace \putrule \let\next=\relax 
        \else \let\next=\putout
            \ifx##1- \advance\count1 by 2 \putspace
            \else    \advance\count0 by 2 \putrule \fi \fi \next}   }
\def\tablespec#1{
    \def\vdimens{\noexpand\tableheight}
    \def\tabby{\tabskip=0pt plus100pt minus100pt}
    \def\r{&################\tabby&\hfil################\unskip}
    \def\c{&################\tabby&\hfil################\unskip\hfil}
    \def\l{&################\tabby&################\unskip\hfil}
    \edef\templ{\noexpand\vdimens ########\unskip  #1 
         \unskip&########\tabskip=0pt&########\cr}
    \tabledelim
    \edef\body##1{ \vbox{
        \tabskip=0pt \offinterlineskip
        \halign to \twidth {\templ ##1}}} }

\newbox\grsign \setbox\grsign=\hbox{$>$}
\newdimen\grdimen \grdimen=\ht\grsign
\newbox\laxbox \newbox\gaxbox
\setbox\gaxbox=\hbox{\raise.5ex\hbox{$>$}\llap
	{\lower.5ex\hbox{$\sim$}}}\ht1=\grdimen\dp1=0pt
\setbox\laxbox=\hbox{\raise.5ex\hbox{$<$}\llap
	{\lower.5ex\hbox{$\sim$}}}\ht2=\grdimen\dp2=0pt
\def\simlt{\mathrel{\copy\laxbox}}

\def\uJy{\ifmmode{\,\mu{\rm Jy}}\else$\,{\mu{\rm Jy}}$\fi}
\def\mJy{\ifmmode{\,{\rm mJy}}\else${\,{\rm mJy}}$\fi}
\def\MHz{\ifmmode{\,{\rm MHz}}\else{$\,{\rm MHz}$}\fi}
\def\GHz{\ifmmode{\,{\rm GHz}}\else{$\,{\rm GHz}$}\fi}
\def\solar{\ifmmode_{\mathord\odot}\else$_{\mathord\odot}$\fi}
\def\Msolar{\ifmmode{\, {\rm M\solar}}\else{${\, {\rm M\solar}}$}\fi}
\def\Rsolar{\ifmmode{\, {\rm R\solar}}\else{${\, {\rm R\solar}}$}\fi}
\def\kms{\ifmmode{\,{\rm km\,s^{-1}}}\else${\,{\rm km\,s^{-1}}}$\fi}
\def\kpc{\ifmmode{\,{\rm kpc}}\else${\,{\rm kpc}}$\fi}
\def\us{\ifmmode{\,\mu{\rm s}}\else$\,{\mu{\rm s}}$\fi}
\def\ms{\ifmmode{\,{\rm ms}}\else$\,{{\rm ms}}$\fi}
\def\y{\ifmmode{\,{\rm y}}\else$\,{\rm y}$\fi}
\def\h{\ifmmode{^{\rm h}}\else$^{\rm h}$\fi}
\def\m{\ifmmode{^{\rm m}}\else$^{\rm m}$\fi}
\def\s{\ifmmode{^{\rm s}}\else$^{\rm s}$\fi}
\def\Lmin{\ifmmode{L_{min}}\else{$L_{min}$}\fi}
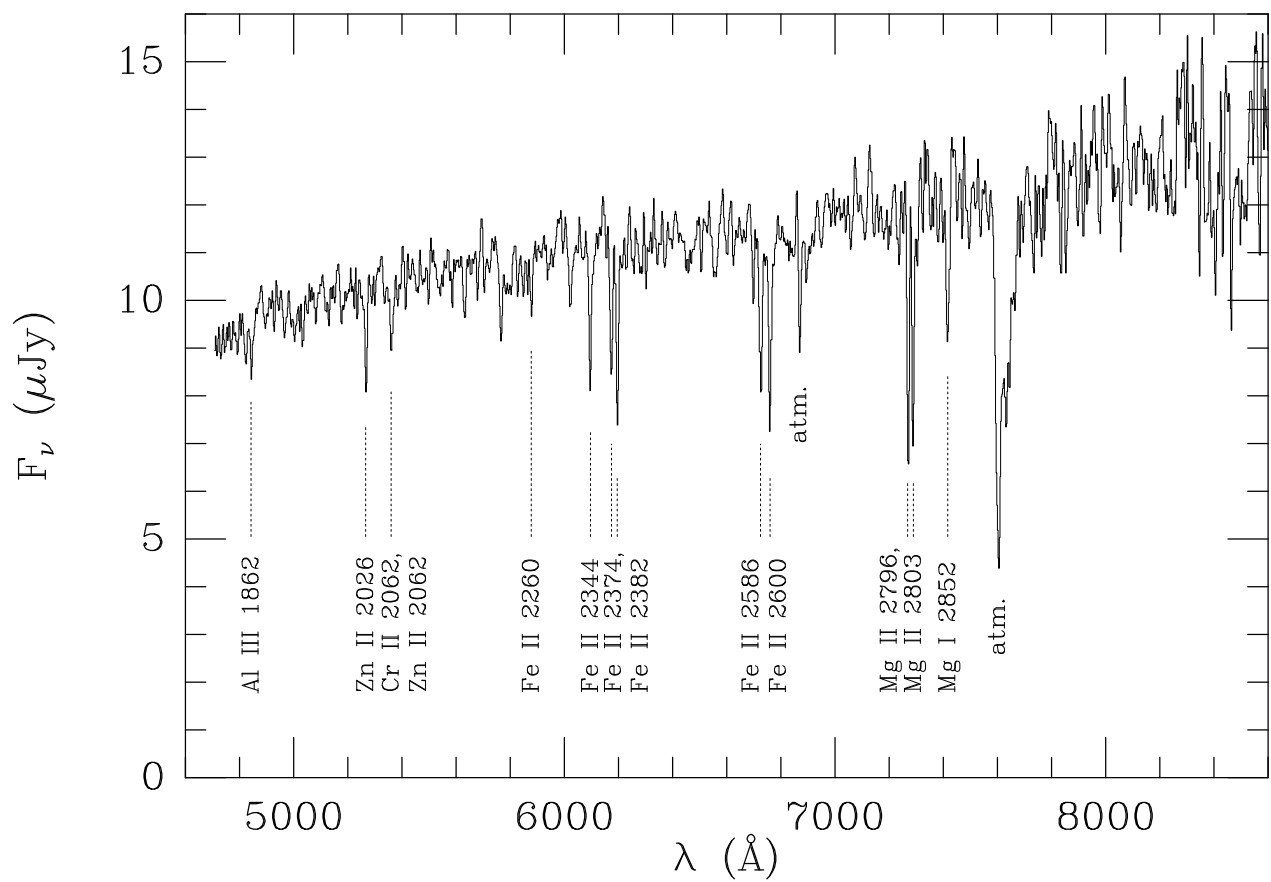
